\documentclass[prc,showpacs,superscriptaddress,twocolumn]{revtex4}

\usepackage{amsmath,bm}
\usepackage{graphicx}
\usepackage{epstopdf}
\usepackage{subfigure}
\usepackage{epsfig}
\usepackage{amsmath,amssymb,amsfonts}
\usepackage{color}
\usepackage[utf8]{inputenc}
\usepackage{verbatim}
\usepackage{array}

\begin{document}

\newcommand{\vk}{{\vec k}}
\newcommand{\vK}{{\vec K}}
\newcommand{\vb}{{\vec b}}
\newcommand{{\vp}}{{\vec p}}
\newcommand{{\vq}}{{\vec q}}
\newcommand{\vQ}{{\vec Q}}
\newcommand{\vx}{{\vec x}}
\newcommand{\dd}{\textrm{\,d}}
\newcommand{\beq}{\begin{equation}}
\newcommand{\eeq}{\end{equation}}
\newcommand{\half}{{\textstyle \frac{1}{2}}}
\newcommand{\gton}{\stackrel{>}{\sim}}
\newcommand{\lton}{\mathrel{\lower.9ex \hbox{$\stackrel{\displaystyle<}{\sim}$}}}
\newcommand{\ee}{\end{equation}}
\newcommand{\ben}{\begin{enumerate}}
\newcommand{\een}{\end{enumerate}}
\newcommand{\bit}{\begin{itemize}}
\newcommand{\eit}{\end{itemize}}
\newcommand{\bc}{\begin{center}}
\newcommand{\ec}{\end{center}}
\newcommand{\bea}{\begin{eqnarray}}
\newcommand{\eea}{\end{eqnarray}}

\newcommand{\beqar}{\begin{eqnarray}}
\newcommand{\eeqar}[1]{\label{#1} \end{eqnarray}}
\newcommand{\pleft}{\stackrel{\leftarrow}{\partial}}
\newcommand{\pright}{\stackrel{\rightarrow}{\partial}}

\newcommand{\eq}[1]{Eq.~(\ref{#1})}
\newcommand{\fig}[1]{Fig.~\ref{#1}}
\newcommand{\eff}{ef\!f}
\newcommand{\alphas}{\alpha_s}

\renewcommand{\topfraction}{0.85}
\renewcommand{\textfraction}{0.1}

\renewcommand{\floatpagefraction}{0.75}
\newcommand{\pp}{p\kern-0.05em p}
\newcommand{\ppbar}{\mathrm{p}\kern-0.05em \overline{\mathrm{p}}}
\newcommand{\pPb}{\ensuremath{\mbox{p--Pb}}}
\newcommand{\PbPb}{\ensuremath{\mbox{Pb--Pb}}}
\newcommand{\GeV}{\ensuremath{\mathrm{GeV}\kern-0.05em}}
\newcommand{\GeVc}{\ensuremath{\mathrm{GeV}\kern-0.05em/\kern-0.02em c}}
\newcommand{\sqrts}{\ensuremath{\sqrt{s_{\mathrm{NN}}}}}
\newcommand{\pT}{\ensuremath{p_{\mathrm{T}}}}
\newcommand{\RL}{\ensuremath{R_{\mathrm{L}}}}
\newcommand{\pTi}{\ensuremath{p_{\mathrm{T},i}}}
\newcommand{\pTtrack}{\ensuremath{p_{\mathrm{T,track}}}}
\newcommand{\sigmaeec}{\ensuremath{\Sigma_{\mathrm{EEC}}}}

% Jet
\newcommand{\kT}{\ensuremath{k_{\mathrm{T}}}}
\newcommand{\pThard}{\ensuremath{p_{\mathrm{T,hard}}}}
\newcommand{\etajet}{\ensuremath{\eta_{\mathrm{jet}}}}
\newcommand{\pTjet}{\ensuremath{p_{\mathrm{T}}^{\mathrm{jet}}}}
\newcommand{\pTchjet}{\ensuremath{p_{\mathrm{T}}^{\mathrm{ch\; jet}}}}
\newcommand{\pTfulljet}{\ensuremath{p_{\mathrm{T}}^{\mathrm{full\; jet}}}}
\newcommand{\pTtruth}{\ensuremath{p_{\mathrm{T,truth}}^{\mathrm{ch\; jet}}}}
\newcommand{\pTdet}{\ensuremath{p_{\mathrm{T,det}}^{\mathrm{ch\; jet}}}}
\newcommand{\Nevent}{\ensuremath{N_{\mathrm{event}}}}
\newcommand{\Ninc}{\ensuremath{N_{\mathrm{jets}}}}
\newcommand{\Sinc}{\ensuremath{\sigma_{\mathrm{jets}}}}

\title{Study of the EEC discrimination power on quark and gluon jet quenching effects in heavy-ion collisions at  $\sqrt{s}=5.02$~TeV}

\date{\today  \hspace{1ex}}

\author{Shi-Yong Chen}
\affiliation{Huanggang Normal University, Huanggang 438000, China}
\affiliation{Key Laboratory of Quark \& Lepton Physics (MOE) and Institute of Particle Physics, Central China Normal University, Wuhan 430079, China}

\author{Zi-Xuan Xu}
\affiliation{Key Laboratory of Quark \& Lepton Physics (MOE) and Institute of Particle Physics, Central China Normal University, Wuhan 430079, China}

\author{Ke-Ming Shen}
\affiliation{East China University of Technology, Nanchang 330013, China}

\author{Wei Dai\footnote{weidai@mail.cug.edu.cn}}
\affiliation{School of Mathematics and Physics, China University of Geosciences, Wuhan 430074, China}

\author{Ben-Wei Zhang}
\affiliation{Key Laboratory of Quark \& Lepton Physics (MOE) and Institute of Particle Physics, Central China Normal University, Wuhan 430079, China}

\author{Enke Wang}
\affiliation{Guangdong Provincial Key Laboratory of Nuclear Science, Institute of Quantum Matter, South China Normal University, Guangzhou 510006, China}
\affiliation{Key Laboratory of Quark \& Lepton Physics (MOE) and Institute of Particle Physics, Central China Normal University, Wuhan 430079, China}

\begin{abstract}

We present a systematic investigation of flavor-dependent jet quenching using energy-energy correlators (EEC) in $\sqrt{\rm s}=5.02$ TeV Pb+Pb collisions. Employing the improved SHELL model, which incorporates collisional and radiative energy loss, as well as medium response, we quantify distinct quenching signatures for quark and gluon jets. Key findings include: (1) Pure quark jets exhibit strong EEC enhancement at large angular scales, while gluon jets show a bimodal enhancement pattern at both small and large scales; (2) Dual-shift decomposition in the EEC ratio reveals shifts toward large primarily driven by energy loss, while small-\RL~shifts extend beyond selection bias and indicate intrinsic enhancement of the gluon-initiated jets; (3) Quark jets experience global suppression of averaged energy weight $\langle\mathrm{weight}\rangle(\RL)$, whereas gluon jets exhibit concentration toward small \RL; (4) Mechanism decomposition identifies elastic energy loss concentrating $\langle\mathrm{weight}\rangle(\RL)$ toward small \RL, radiative loss dominating quark jet modification, and medium response amplifying large \RL ~enhancement via soft hadrons. The observed flavor dependence in EEC modifications is dominantly driven by intrinsic jet structure differences rather than medium-induced mechanisms. We propose photon-tagged jets as quark proxies and inclusive charged-hadron jets as gluon proxies, finding they reproduce the respective flavor-specific quenching patterns. Our work establishes the EEC as a precision probe of color-charge-dependent jet-medium interactions, providing new constraints for the detailed $\hat{q}$ extraction and QGP tomography, while highlighting the critical role of pre-quenching flavor asymmetries.

\end{abstract}


\maketitle

%%%introduction%%%%
%%
\section{Introduction}\label{sec:1}

The creation of a deconfined quark-gluon plasma (QGP) in relativistic heavy-ion collisions provides a unique opportunity to study quantum chromodynamics (QCD) under extreme conditions in the laboratory~\cite{Gyulassy:1990ye,Luo:2017faz,Tang:2020ame,Zhang:2021xib,Shou:2024uga}. 
Highly energetic jets-collimated sprays of particles resulting from the fragmentation of hard-scattered partons produced in the initial collisions, served as precision probes to the novel properties of this QCD medium. 
The phenomenon of jet quenching, characterized by the significant suppression of high-transverse momentum (\pT) hadrons and jets relative to scaled proton-proton collisions, arises primarily from parton energy loss mechanisms, such as medium-induced gluon radiation and elastic collisions~\cite{Wang:1992qdg,Gyulassy:2003mc,Qin:2015srf}. 
Quantifying this energy loss provides critical, direct insights into the QGP's transport coefficients. 
To dissect the complex space-time evolution of the in-medium parton shower and differentiate between competing energy loss mechanisms, jet substructure observables have emerged as indispensable, high-resolution tools~\cite{Young:2011qx,He:2011pd,Neufeld:2010fj,Zapp:2012ak, Dai:2012am, Ma:2013pha, Senzel:2013dta, Casalderrey-Solana:2014bpa,Milhano:2015mng,Chang:2016gjp,Majumder:2014gda, Chen:2016cof, Chien:2016led, Apolinario:2017qay,Connors:2017ptx,Zhang:2018urd,Dai:2018mhw,Luo:2018pto,Chang:2019sae,Wang:2019xey,Chen:2019gqo,Chen:2020kex,Wang:2020qwe,Yan:2020zrz,Wang:2020ukj,Zhang:2021sua,Li:2024uzk}. 
These observables quantify the modification of the jet's internal energy flow and fragmentation pattern caused by interactions with the QGP. 
Recent significant advances include detailed measurements of differential jet shapes~\cite{Chien:2015hda,CMS:2018jco,Luo:2018pto,Chang:2019sae,Yang:2022nei}, fragmentation functions~\cite{Casalderrey-Solana:2016jvj,CMS:2018mqn,ATLAS:2018bvp,Chen:2020tbl}, groomed jet observables~\cite{Chien:2016led,ALargeIonColliderExperiment:2021mqf}, collectively offering a multi-dimensional tomography of jet-medium interactions.

Among jet substructure observables, the energy-energy correlator (EEC)~\cite{Basham:1977iq,Basham:1978bw,Basham:1978zq,PLUTO:1985yzc} is particularly significant due to its sensitivity to soft and collinear radiation patterns within jets. Defined as the energy-weighted angular correlation between particle pairs, the EEC probes jet internal structure in elementary collisions and has emerged as a powerful tool for studying QCD dynamics~\cite{Larkoski:2013eya,Larkoski:2015uaa,Dixon:2018qgp,Chen:2019bpb,Henn:2019gkr,Chen:2020vvp,Gao:2020vyx,Li:2021txc,Komiske:2022enw,Neill:2022lqx,Liu:2022wop,Chen:2022swd,Holguin:2022epo,Lee:2022uwt,Liu:2023aqb,Cao:2023oef,Chen:2023zlx,Jaarsma:2023ell,Lee:2023npz,Kang:2023big,Kang:2023gvg,Chen:2024nfl,Barata:2023zqg,Holguin:2023bjf,Barata:2024nqo,Alipour-fard:2025dvp,Craft:2022kdo}. In heavy-ion collisions, it elucidates the onset of color coherence in medium-induced splittings~\cite{Andres:2022ovj,Andres:2023xwr,Barata:2023vnl,Barata:2023bhh,Singh:2024vwb,Andres:2023ymw,Andres:2024xvk,Andres:2024ksi}, reveals the medium response effect, mass hierarchy in heavy-flavor jets~\cite{Yang:2023dwc,Bossi:2024qho,Xing:2024yrb,Barata:2024ieg}, and exposes cold nuclear matter effects in small systems~\cite{Devereaux:2023vjz,Fu:2024pic,Barata:2024wsu}. The impact of selection bias on the modification of the EEC is also discussed in~\cite{Andres:2024hdd,Andres:2024pyz}. Moreover, EEC has also garnered significant attention among experimentalists~\cite{Tamis:2023guc,CMS:2024ovv,ALICE:2024dfl,ALICE:2025igw}, with STAR, CMS, and ALICE collaborations conducting dedicated measurements of this observable in heavy-ion collisions. 

Flavor-dependent parton energy loss is a fundamental aspect governing jet evolution within the QGP~\cite{Wang:1998bha,Gyulassy:2003mc}. Distinct Casimir color factors not only lead to differences in the magnitude of energy loss between quarks and gluons but also subject them to differential in-medium splitting~\cite{Baier:1996sk,Wang:2001ifa,Arnold:2000dr,Gyulassy:2000er}. This inherent difference in energy loss rates manifests directly as distinct medium modifications in jet substructure observables, such as jet charge~\cite{Chen:2019gqo,Li:2019dre,CMS:2020plq} and jet shape~\cite{Chien:2015hda,Chien:2018dfn,CMS:2018jco,Yan:2020zrz}.
However, the in-medium radiation pattern of this quark/gluon difference on jet substructure observable in heavy-ion collisions remains lacking in studies. Therefore, it is compelling to investigate the flavor-dependent jet quenching pattern within the finer in-jet substructure using the EEC.

The remainder of the paper is organized as follows. In Sec .~\ref {sec:2}, we introduce the definition of the EEC used for the ALICE experimental study, and the p+p baseline of the EEC distributions for inclusive jets in three different jet transverse momentum intervals will be calculated to confront the experimental data. In Sec .~\ref {sec:3} we calculate and compare the EEC distributions for inclusive jets in Pb+Pb and p+p at $\sqrt{\rm s} = 5.02$~TeV for the transverse momentum interval $40- 60$~GeV, to derive the medium modification ratios (A+A/p+p). The in-depth phenomenology exploration of such an observable is also presented and helps us discuss the differences in the jet quenching of quark- and gluon-initiated jets from the following four aspects: their respective Energy-Energy Correlation (EEC) distributions, the impacts of different jet quenching mechanisms on them respectively, the different modifications of the fine structures revealed by the intra-jet EEC, and their different selection bias effects. Finally, we close this paper to a summary in Sec .~\ref {sec:4}.

\section{EEC distributions in p+p collisions}\label{sec:2}
The ALICE defined EEC describes the energy-weighted cross-section of particle pairs as functions of the angular distance between each pair as follows~\cite{ALICE:2025igw}:

%\lipsum[1]
\begin{widetext}
\begin{eqnarray}
%\centering
\Sigma_{\text{EEC}}(R_{\text{L}}) = \frac{1}{ N_{\text{jet}}\cdot\Delta R}\int_{R_{\text{L}}-\frac{1}{2}\Delta R}^{R_{\text{L}}+\frac{1}{2}\Delta R}\sum_{\text{jets}}\sum_{i, j}\frac{p_{\text{T}, i}p_{\text{T}, j}}{p^2_{\rm T, jet}}\delta(R_{\text{L}}'-R_{\text{L}, ij})~\dd R_{\text{L}}'~.
\label{EEC-equ}
\end{eqnarray}
\end{widetext}
%\lipsum[1]
where all final state particle pairs ($i$, $j$) inside each jet are summed up. The angular distance between each pair is defined in the $\eta-\varphi$ plane as $R_{\text{L}, ij} = \sqrt{(\varphi_j - \varphi_i)^2 + (\eta_j - \eta_i)^2}$, which $\Delta R$ is the angular bin width and $N_{\text{jet}}$ is the total number of jets.
Then we noticed it is a distribution observable defined in a jet.

\begin{figure}[!htb]
\centering
\includegraphics[width=9.0cm,height=8.5cm]{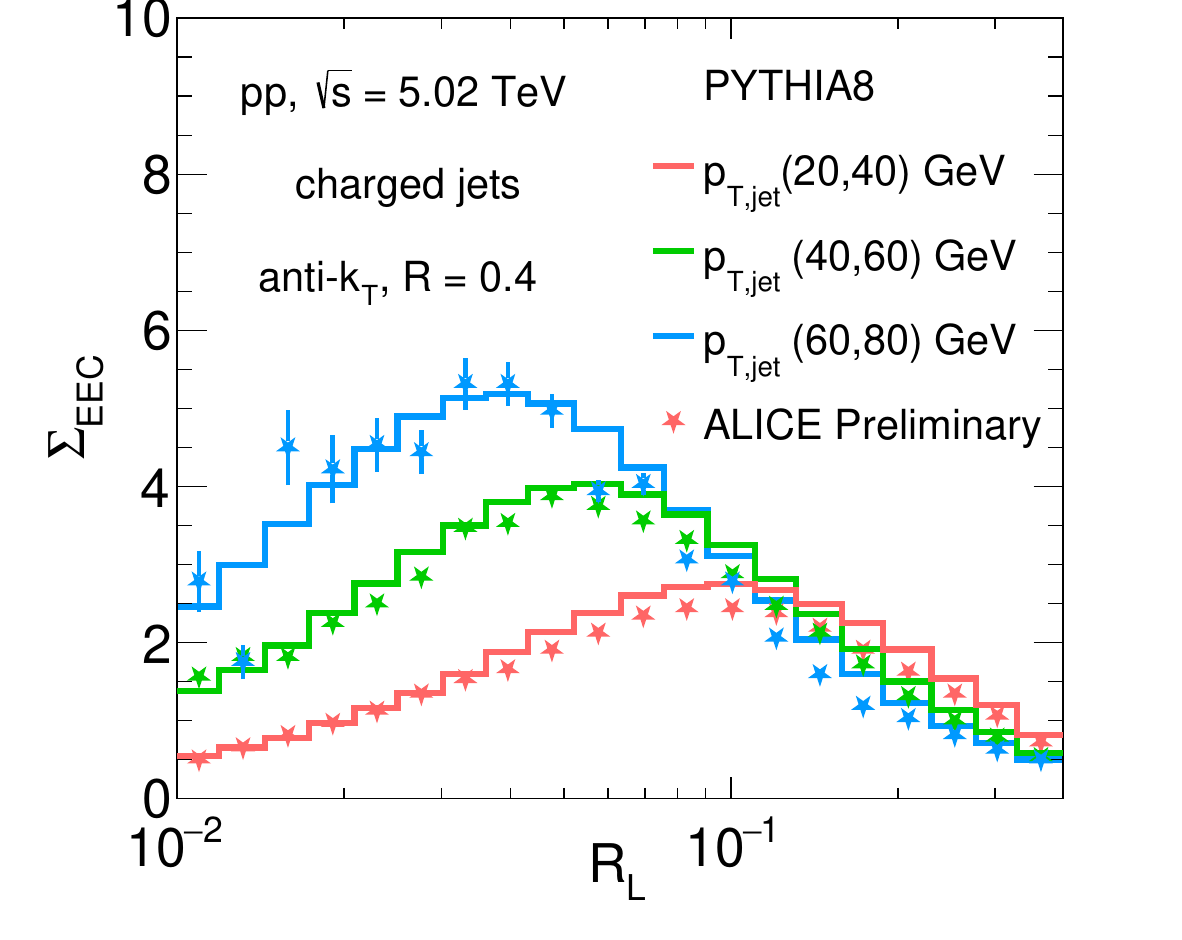}
\caption{PYTHIA8 simulation results of defined $\Sigma_{\text{EEC}}$ distributions as functions of $R_{\rm L}$ for inclusive charged jets with a jet size of $R=0.4$ in three jet transverse momentum intervals: $20 \ {\rm GeV} <p_{T, \rm jet}<40 $~GeV, $40 \ {\rm GeV} <p_{T, \rm jet}<60 $~GeV and $60 \ {\rm GeV} <p_{T, \rm jet}<80 $~GeV produced in $\rm p+p$ collisions at $\sqrt{s}=5.02$~TeV. The results are compared with the ALICE experimental data~\cite{ALICE:2024dfl}.}
\label{fig:baseline}%{}
\end{figure}

We start by calculating the defined EEC distributions of inclusive jets in p+p collisions to provide a baseline for further studies. In this work, we use a Monte Carlo (MC) event generator PYTHIA v8.309~\cite{Sjostrand:2014zea}
with Monash 2013 tune~\cite{Skands:2014pea} to simulate jet productions in p+p collisions.
To confront our calculated results with experimental data, we use the same kinematic cuts of events as adopted by the ALICE measurements~\cite{ALICE:2024dfl}.  All jets are reconstructed by the anti-$k_{\rm T}$ algorithm with radius parameter $R=0.4$ from charged particles with $p_{\rm T}\geq 1$~GeV using the FASTJET v3.4.0 package~\cite{Cacciari:2008gp}. These reconstructed jets are accepted in the transverse momentum range $20 \ {\rm GeV} <p_{\rm T, \rm jet}<80$~GeV and rapidity range of $\left|\eta_{\rm jet}\right|<0.5$. Our numerical results of $\Sigma_{\text{EEC}}$ as functions of \RL~for inclusive charged jets and their comparisons with ALICE data in p+p collisions at $\sqrt{s}= 5.02$~TeV are shown in Fig.~\ref{fig:baseline}.

We can observe that our numerical results show fairly good agreement with experimental measurements in p+p collisions in the three $p_{\rm T}$ intervals, which will serve as input and baseline for the subsequent study in Pb+Pb collisions. 
The EEC distributions are shifted to a lower \RL ~region with the increasing jet $p_{\rm T}$.
We can conclude that, with the increment of jet $p_{\rm T}$, the EEC distributions shift to smaller \RL, and the height of the distribution will increase which is surely affected by the different number of particle pairs within one jet in each $p_{\rm T}$ interval. To extract such an effect, it is natural to divide the EEC observable into two major parts rewritten as:

\begin{eqnarray}
%\centering
\Sigma_{\text{EEC}}(R_{\text{L}}) = \frac{N^{\rm total}_{\rm pair}}{N_{\rm jet}} \thinspace \cdot
\frac{\Delta N_{\rm pair}}{N^{\rm total}_{\rm pair} \Delta R}(R_{\rm L}) \thinspace \cdot \langle \rm weight \rangle(R_{L})
\label{EEC-fac}
\end{eqnarray}
where we average the energy weight term, $p_{\text{T}, i}p_{\text{T}, j}/p^2_{\rm T, jet}$, to every particle pair within every jet in each angular bin $\Delta R$, denoted as $\langle \rm weight \rangle$, therefore we can simply replace the integration and then sum over jets and $i, j$ into a summed number of pairs within each bin $\Delta R$, denoted as $\Delta N_{\rm pair}$. Consequently, Eq.~\ref{EEC-fac} appears as: the averaged number of particles within a jet, normalized \RL~distribution on the number of pairs, and the averaged energy weight distribution as a function of \RL. It will benefit further explorations and discussions.

\begin{figure}[!htb]
\centering
\includegraphics[scale=0.45]{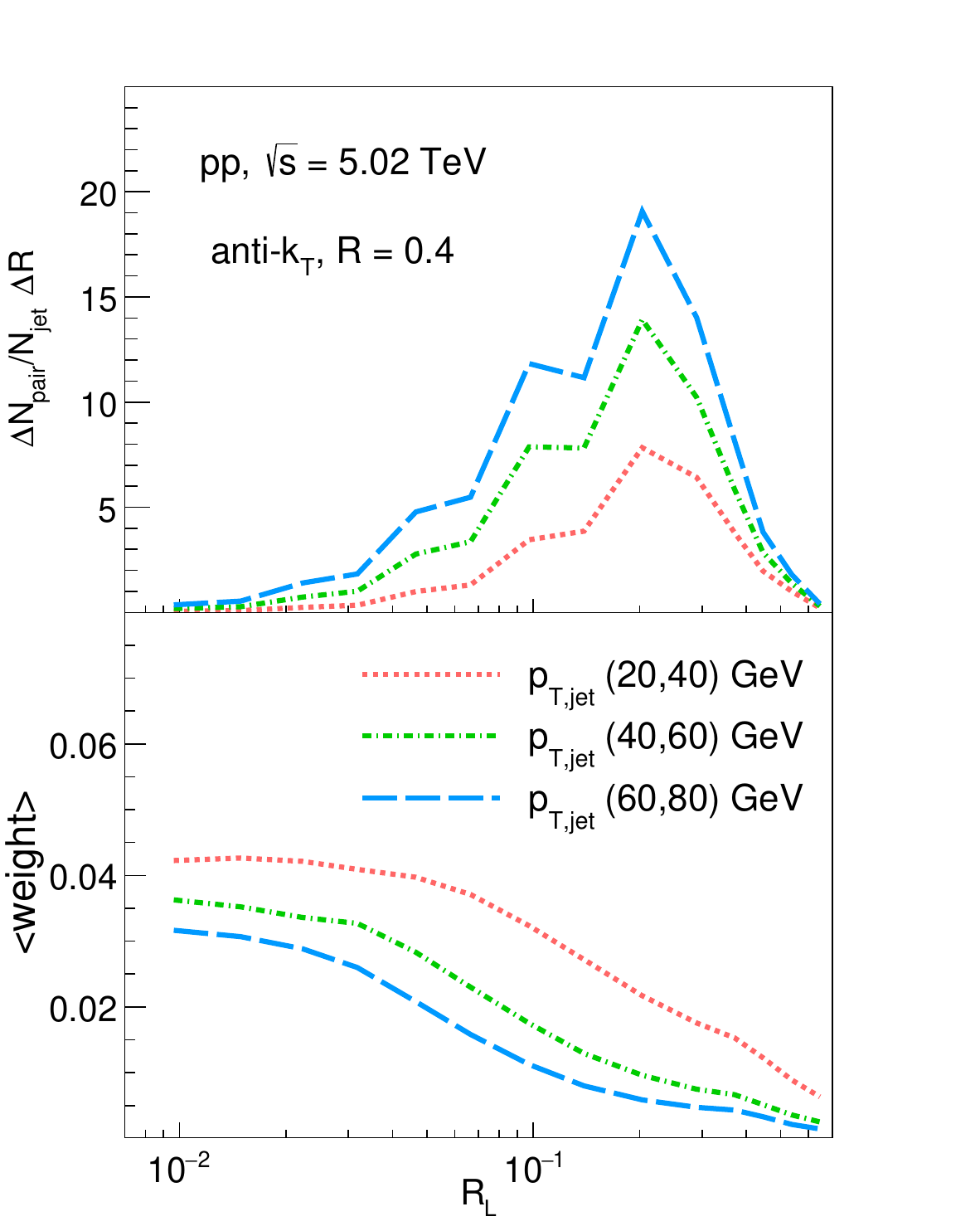}
\caption{Number of jets normalized pair-number distributions (upper panel) and averaged energy weight (lower panel) distributions as functions of $R_{L}$ for inclusive charged jets in three jet transverse momentum intervals: $20 \ {\rm GeV} <p_{T, \rm jet}<40 $~GeV, $40 \ {\rm GeV} <p_{T, \rm jet}<60 $~GeV and $60 \ {\rm GeV} <p_{T, \rm jet}<80 $~GeV produced in $\rm p+p$ collisions at $\sqrt{s}=5.02$~TeV.} 
\label{fig:ppangle}%{}
\end{figure}

We plot the \RL~distributions of the total number of pairs normalized to every inclusive charged jet produced in $\rm p+p$ collisions at $\sqrt{s}=5.02$~TeV for all three $p_{\rm T,jet}$ intervals in the upper panel of Fig.~\ref{fig:ppangle}, denoted as the product of the first two terms in Eq.~\ref {EEC-fac}. The corresponding \RL~distributions of the averaged weight $\langle \rm weight \rangle$ are also demonstrated in the bottom panel of Fig.~\ref{fig:ppangle}. 

We find in the upper panel that most of the particle-pairs are distributed around $R_{\rm L}=0.2$, and there is no jet transverse momentum dependence of the peak position. The distribution height increases with the enhancement of $p_{\rm T,jet}$ because the number of constituents within the jets increases as a consequence. In the bottom panel, the role of the $\langle \rm weight \rangle$ is to shift the distribution of particle pairs toward smaller \RL~directions, while the heights of the $\langle \rm weight \rangle$ distribution are also determined by the jet transverse momentum. Higher jet transverse momentum leads to more constituent particles within the jet, consequently resulting in lower $\langle \rm weight \rangle$ values. 

\section{EEC distributions in A+A collisions}\label{sec:3}
\subsection{Improved SHELL model}\label{sec:31}
In relativistic heavy-ion collisions, partons produced from initial hard scatterings undergo energy loss through interactions with the QGP. To simulate this process, we employ the improved SHELL model—a Monte Carlo framework that concurrently handles both elastic, inelastic scattering processes and the medium response during jet propagation. The SHELL model has been quantitatively validated against multiple experimental observables, establishing its reliability for jet quenching studies~\cite{Yan:2020zrz,Chen:2022kic,Dai:2022sjk,Li:2022tcr,Wang:2024plm,Li:2024pfi}. The model initializes parton positions via sampling from a Glauber-model-based nuclear geometry~\cite{Alver:2008aq}, then they will transport through the QGP step-by-step. The radiative energy loss mechanism is implemented through a stochastic implementation approach. The probability of a gluon radiation when traversing through the QGP medium during each time step $\Delta t$ can be expressed as:
\begin{eqnarray}
P_{rad}(t,\Delta t)=1-e^{-\left\langle N(t,\Delta t)\right\rangle} \, .
\label{eq:g4}
\end{eqnarray}
Here $\left\langle N(t,\Delta t)\right\rangle$ is the averaged number of radiated gluons, which can be calculated
from the medium induced radiated gluon spectrum within the Higher-Twist (HT) method\cite{Majumder:2009ge,Guo:2000nz,Zhang:2003yn,Cao:2017hhk}:
\begin{eqnarray}
\frac{\dd N}{\dd x \dd k^{2}_{\perp}\dd t}=\frac{2\alpha_{s}C_sP(x)\hat{q}}{\pi k^{4}_{\perp}}\sin^2(\frac{t-t_i}{2\tau_f})(\frac{k^2_{\perp}}{k^2_{\perp}+x^2M^2})^4
\label{eq:g5}
\end{eqnarray}
where $\alpha_{s}$ is the strong coupling constant, $x$ and $k_\perp$ devote the energy fraction and the $p_{T}$ of the radiated gluon, $M$ is the mass of the parent parton. Only the gluon with a lower $x_{min}=\mu_{D}/E$ cut-off is allowed to emit, and $\mu_{D}$ is the Debye screening mass. $P(x)$ is the QCD splitting function in vacuum, $C_s$ devotes the Casimir factor for gluon ($C_A$) and quark ($C_F$). $\tau_f=2Ex(1-x)/(k^2_\perp+x^2M^2)$ is the formation time of the radiated gluons. $\hat{q}=q_{0}(T/T_{0})^{3}p_{\mu}u^{\mu}$ is the jet transport parameter, where $q_{0}=1.5~GeV^{2}/fm$, $u^{\mu}$ is the local velocity of the QGP, and $T_{0}$ is the initial temperature. The jet transport parameter is used to control the magnitude of energy loss due to jet-medium interaction.
The number of radiated gluons is sampled from a Poisson distribution during each inelastic scattering. 
\begin{eqnarray}
P(n_{g},t,\Delta t)=\frac{\left\langle N(t,\Delta t)\right\rangle^{n_{g}}}{n_{g}!}e^{-\left\langle N(t,\Delta t)\right\rangle} \,
\label{eq:g4}
\end{eqnarray}
In our calculation, $P_{rad}(t,\Delta t)$ would first be evaluated to determine whether the radiation happens during $\Delta t$. If accepted, the Possion distribution $P(n_{g},t,\Delta t)$ is used to sample the number of radiated gluons. At last, the energy fraction ($x$) and transverse momentum ($k_{\perp}$) of the radiated gluon could be
sampled based on the spectrum shown in Eq.~\ref {eq:g5}.

To calculate the collisional energy loss of these showered partons, a Hard Thermal Loop (HTL) formula\cite{Neufeld:2010xi} has been adopted in this work:
$\frac{\dd E^{coll}}{\dd t}=\frac{\alpha_{s}C_{s}\mu^{2}_{D}}{2}\ln\frac{\sqrt{ET}}{\mu_{D}}$. The space-time evolution of the expanding fireball is given by the CLVisc hydrodynamic model\cite{Shen:2014vra}. When local temperature falls below $T_c=165$~MeV, all the showered partons stop their propagation in the QGP medium and fragment into hadrons.
In this work, we first construct strings using the colorless method developed by the JETSCAPE collaboration~\cite{Putschke:2019yrg}, then perform hadronization and hadron decays using the PYTHIA Lund string method.

We also include the medium response effect in the calculation by considering that the lost energy in the collisional process is deposited into the evolved QGP medium. This deposited energy disturbs the medium, exciting a hydrodynamic wake correlated with the parton's direction. Following freeze-out, this wake hadronizes, producing soft hadrons that can be reconstructed within the jet cone. To incorporate this medium response effect, we employ a hybrid approach based on the Cooper-Frye freeze-out prescription with perturbations~\cite{Cooper:1974mv,Casalderrey-Solana:2016jvj}. The resulting distribution of wake particles is given by:
\begin{align}
  \label{eq:wake}
  E\frac{\dd \Delta N}{\dd^3p}=&\frac{1}{32 \pi} \, \frac{m_{\rm T}}{T^5} \, \cosh(y-y_j) \exp\left[-\frac{m_{\rm T}}{T}\cosh(y-y_j)\right] \notag \\
      &\times \Big\{ p_T \Delta P_{\rm T} \cos (\phi-\phi_j) \notag \\
      &+\frac{1}{3}m_{\rm T} \, \Delta M_{\rm T} \, \cosh (y-y_j) \Big\},
\end{align}
Here, $m_{\rm T}$, $p_{\rm T}$, $y$, and $\phi$ denote the transverse mass, transverse momentum, rapidity, and azimuthal angle of the emitted wake hadrons, respectively. The variables $y_j$ and $\phi_j$ represent the rapidity and azimuthal angle of the initiating energetic parton. $T$ denotes the freeze-out temperature of the hot QCD medium. $\Delta M_{\rm T} = \Delta E / \cosh(y_j)$ and $\Delta P_{\rm T}$ are the transverse mass and transverse momentum transferred from the jet to the medium, where $\Delta E$ is the deposited energy.

\subsection{Collisional vs. Radiative vs. Medium Response}\label{sec:32}
\begin{figure}[!htb]
\centering
\includegraphics[scale=0.45]{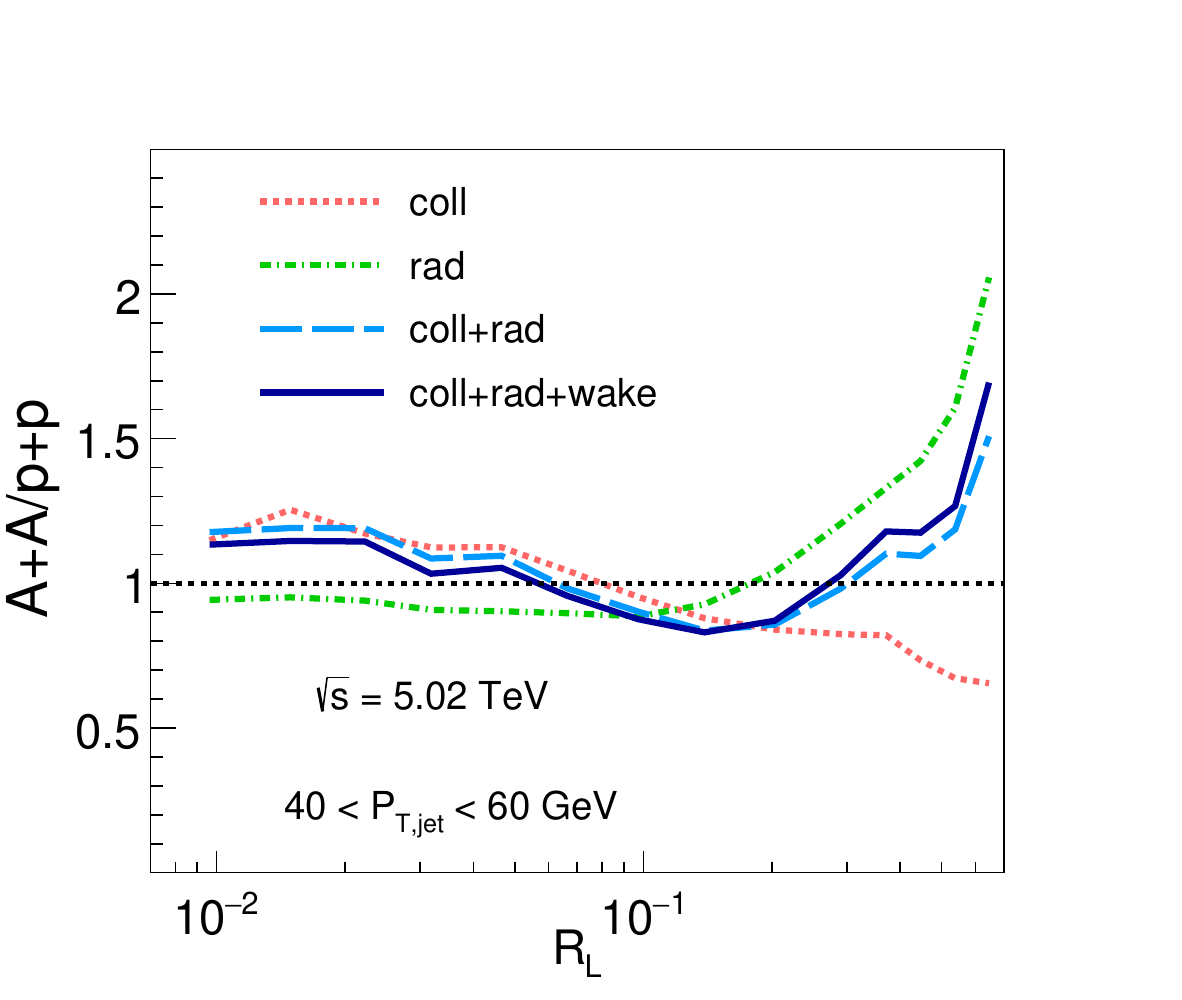}
%\vspace{-1.5cm}
\caption{ A+A/p+p ratios of the $\Sigma_{\text{EEC}}$ distributions as functions of \RL~ for inclusive charged jets with a radius of $R=0.4$ in central ($0-10\%$) $\rm Pb+Pb$ collisions at $\sqrt{\rm s}=5.02$~TeV in jet $p_{\rm T}$ interval $40-60$~GeV calculated considering 4 scenarios of jet quenching mechanism: collisional energy loss only, radiative energy loss only, collisional+radiative energy loss and collisional+radiative energy loss+medium response.}
\label{fig:aappall}%{}
\end{figure}

The angular distribution of the EEC within a jet inherently reflects medium-induced modifications of jet constituents, offering a direct connection to the scale and structure of the QGP. This distribution can be used to study jet-induced medium response, medium-induced radiation, and transverse momentum broadening. We calculate the $\Sigma_{\text{EEC}}$ distributions as functions of \RL~ for inclusive charged jets in Pb+Pb collisions at $\sqrt{\rm s}=5.02$~TeV to predict the possible ALICE measurement. We demonstrate in Fig.~\ref{fig:aappall} the A+A/p+p ratio of the $\Sigma_{\text{EEC}}$ distributions as functions of \RL~with the jet radius of $R=0.4$ in the jet $p_{\rm T}$ interval of $40-60$~GeV shown in the solid blue line. We find clear enhancements at $R_{\rm L}>0.3$ and $R_{\rm L}<0.06$, as well as suppression at $ R _ {\rm L} $ around $0.06-0.3$. From the distribution shifting point of view, it implies there are shifts toward larger \RL~and smaller \RL~at the same time. There have to be two effects competing with each other. To isolate the contributions and effects from different jet quenching mechanisms, we also plot in the same figure the A+A/p+p ratios as functions of \RL: only considering radiative energy loss, denoted as the dashed dotted line; only considering collisional energy loss, denoted as the dotted line; considering both collisional and radiative energy loss but excluding medium response effect, denoted as the dashed line. We will apply the same notation in the following discussion.

Firstly, we find that the pure collisional energy loss process leads to a gradual enhancement in the smaller \RL~region and a gradual suppression in the larger \RL~region, indicating that the energy loss mechanism, which does not alter the particle propagation direction, results in a concentration of the energy weight distribution toward smaller \RL~values. Secondly, the pure radiative energy loss process leads to a rapid enhancement with increasing \RL~in $R_{\rm L}>0.2$ and a slight suppression at $R_{\rm L}<0.2$, indicating a clear broadening effect toward larger \RL~values in the distribution. It is a typical consequence of the medium-induced gluon radiation. The final A+A/p+p ratio is almost exactly the result of the competition of the two mechanisms. The medium response effect seems visible and mild in this specific observable, but it has the same diffusion behaviour of enhancing the distribution in the larger \RL~values and suppressing that in the smaller \RL~region as the radiative energy loss process. 

To gain deeper insight into the physical implications of this result, we examine the theoretical predictions from three perspectives: probing finer jet substructures revealed by the EEC observable itself, specifically the distribution of average paired-energy weight versus \RL~shown in Fig.~\ref{fig:ppangle}; impact of selection bias, which is a common effect in jet substructure observables, on the EEC measurements; capability of this observable to distinguish quark-initiated jets from gluon-initiated jets in jet quenching, along with the underlying mechanisms for such differences.

\subsection{In-jet finer substructure as revealed by the EEC}\label{sec:33}

\begin{figure}[!htb]
\centering
\includegraphics[scale=0.45]{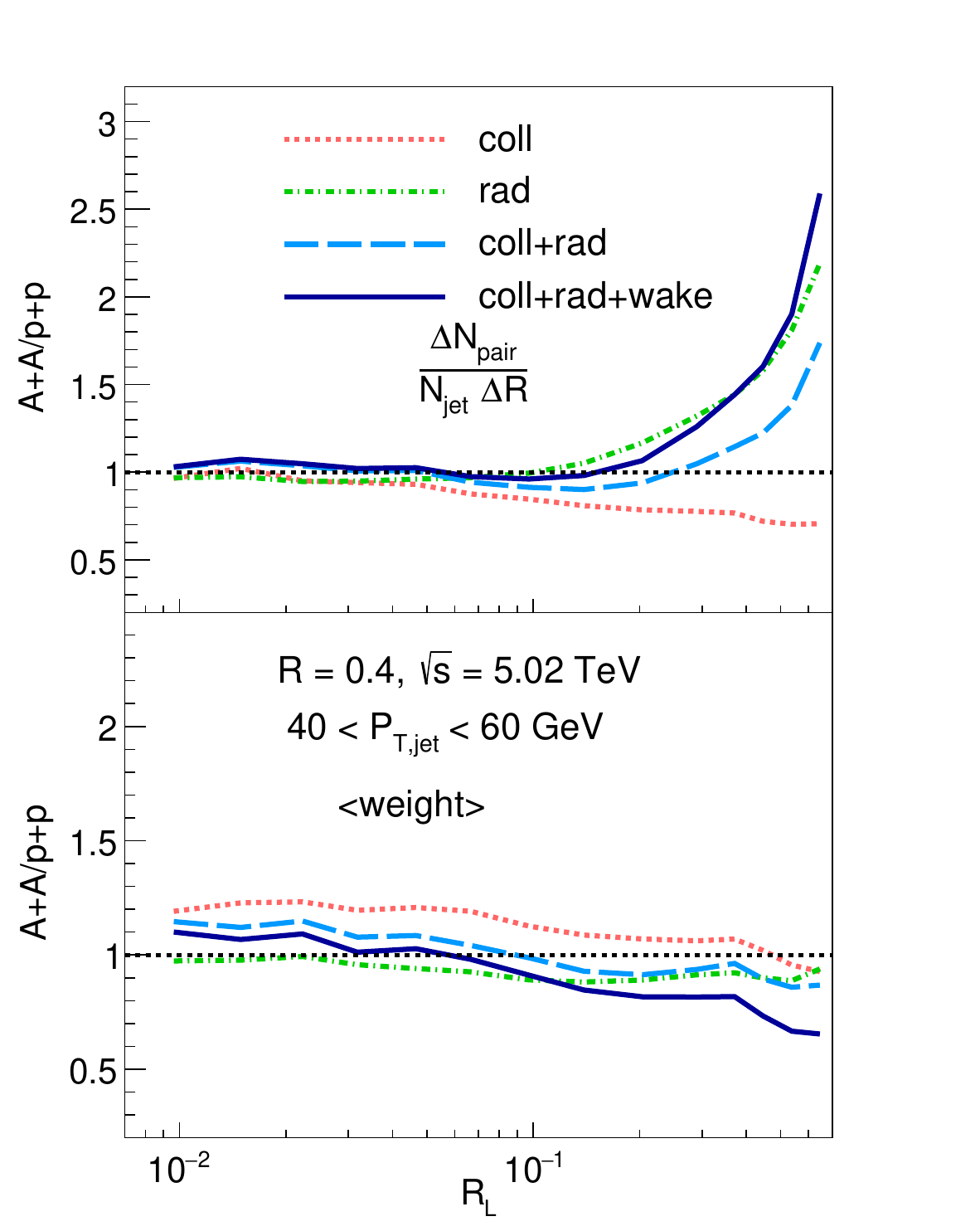}
\caption{
Ratio of the \RL~distribution of paired-particle counts normalized to the number of jets in $\rm Pb+Pb$ collisions to that in $\rm p+p$ collisions (upper panel) and the ratio the \RL-distribution of averaged energy-energy weight in $\rm Pb+Pb$ collisions to that in $\rm p+p$ collisions (lower panel) for inclusive charged jets with the size of $R=0.4$ in the jet transverse momentum interval of $40 \ {\rm GeV} <p_{T, \rm jet}<60 $~GeV produced at $\sqrt{s}=5.02$~TeV. The results of the four jet quenching scenarios are also plotted same way as in Fig.~\ref{fig:aappall}.}
\label{fig:angle-r}%{}
\end{figure}

Applying the strategy discussed in Eq.~\ref{EEC-fac} and Fig.~\ref{fig:ppangle}, we firstly calculate the ratio of the \RL-distribution of paired-particle counts normalized to the number of jets in $\rm Pb+Pb$ collisions to that in $\rm p+p$ collisions shown in the upper panel of Fig.~\ref{fig:angle-r}. Elastic energy loss leads to a reduction in the \RL-distribution of paired-energy counts in A+A compared to the p+p case, and this reduction becomes more pronounced with increasing \RL. The overall suppression demonstrates a reduction in the number of particle pairs. The slope of the curve indicates that it leads to an enhancement of the distribution in the small \RL ~region relative to that in the large \RL ~region. The dashed-dotted line demonstrates that radiative energy loss causes the \RL-distribution of paired-energy counts for jets in A+A collisions to shift towards larger \RL. The dashed line shows how the two mechanisms compete with each other. The comparison between the total effect represented by the solid line and the dashed line reveals that the effect of medium response on the \RL-distribution of paired-particle counts is manifested as an enhancement in the large \RL~region. Furthermore, the medium response resulting in an increase in the number of jet constituents contributes entirely to the distribution in this specific region.
%for inclusive charged jets with the size of $R=0.4$ in the jet transverse momentum interval of $40 \ {\rm GeV} <p_{T, \rm jet}<60 $~GeV produced at $\sqrt{s}=5.02$~TeV 

From the bottom panel of Fig.~\ref{fig:angle-r}, we observe that all types of jet quenching mechanisms lead to an elevation of the averaged energy-energy weights ($\langle \rm weight \rangle$s) in the small \RL~region relative to those in the large \RL~region. This behavior explains why the A+A/p+p ratio of the EEC shown in Fig.~\ref{fig:aappall} exhibits an enhancement in the small \RL~region, while its rise, when moving towards the large \RL~region, is weaker than that observed in the paired-particle count distribution in the upper panel. The ratio value obtained from radiative energy loss is lower than that from collisional energy loss because radiative energy loss increases the number of particle pairs, whereas collisional energy loss reduces it. The inclusion of the medium response effect, due to the resulting increase in the number of jet constituents, causes a reduction in the $\langle \rm weight \rangle$s. This creates a stark contrast to the sharp rise it induces in the paired-particle count distribution within the large \RL ~region, leading to a mild overall effect of the medium response to the A+A/p+p ratio of EEC. Therefore, we find that the upper and lower parts shown in the figure exhibit exactly opposite trends in their response to changes in the number of jet constituents, resulting in a partial cancellation effect.

\subsection{Quark-initiated jets versus  gluon-initiated jets}\label{sec:34}

Still, we need to examine the reason why the $\langle \rm weight \rangle$ distribution in A+A collisions is relatively more concentrated towards the small \RL~region. As suggested by the jet-$p_{\rm T}$ dependence of the EEC distribution shown in Fig.~\ref{fig:baseline}, this phenomenon may be attributed to selection bias. But does this imply that energy loss mechanisms cannot drive the $\langle\mathrm{weight}\rangle$ towards larger \RL~angles? Is this phenomenon consistent for both quark- and gluon-initiated jets? We find that a comprehensive understanding of flavor-dependent jet-quenching effects on the EEC observable remains elusive. To address this, we intend to conduct further discussion focusing on the following aspects: initial p+p production, different jet quenching mechanisms, in-jet finer substructure as revealed by the EEC, selection bias effect, and flavor-dependent attribution. We begin our discussion by comparing purely quark-initiated jets and purely gluon-initiated jets generated by PYTHIA.

\textcolor{blue}{{\bf \textit{initial p+p production}}~-~} Given the experimental challenge of identifying or defining quark versus gluon-initiated jets, we adopt a simulation-based approach. Using PYTHIA 8, we generate events from hard processes constrained to produce exclusively quarks or exclusively gluons. Jets are then reconstructed in these events and used for our case study. 

\begin{figure}[!htb]
\centering
\includegraphics[scale=0.45]{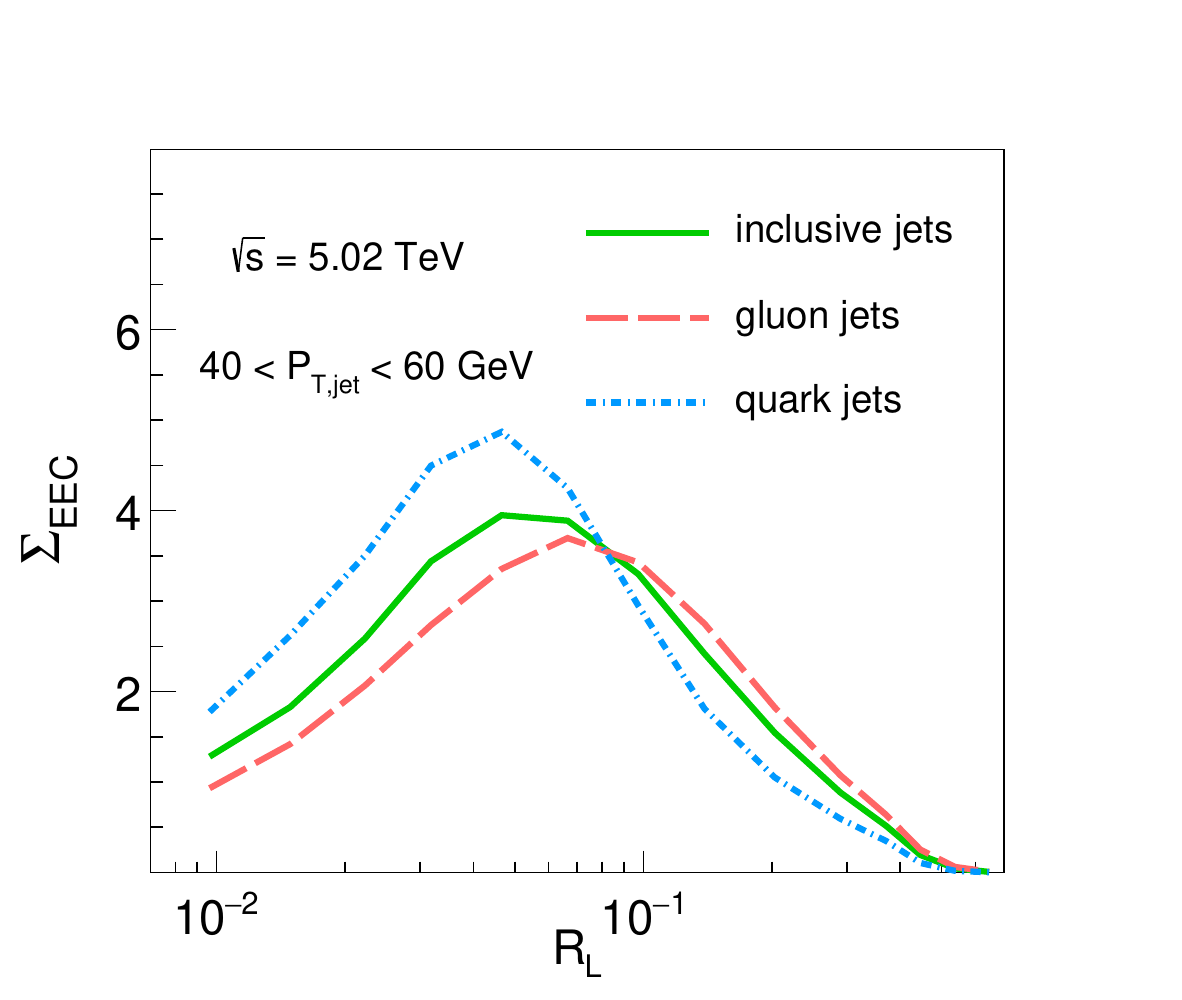}
\caption{PYTHIA8 simulation results of defined $\Sigma_{\text{EEC}}$ distributions as functions of $R_{\rm L}$ for inclusive charged jets, pure gluon-initiated jets and quark -initiated jets with a jet size of $R=0.4$ in the jet transverse momentum interval of $40 \ {\rm GeV} <p_{T, \rm jet}<60 $~GeV produced in $\rm p+p$ collisions at $\sqrt{s}=5.02$~TeV.}
\label{fig:quarkgluon}%{}
\end{figure}

In Fig.~\ref{fig:quarkgluon} we plot $\Sigma_{\text{EEC}}$ distributions as functions of $R_{\rm L}$ for inclusive charged jets, pure gluon-initiated and quark -initiated jets with a size of $R=0.4$ in the jet transverse momentum interval of $40 \ {\rm GeV} <p_{\rm T,jet}<60 $~GeV produced in $\rm p+p$ collisions at $\sqrt{s}=5.02$~TeV simultaneously. We observe that gluon-initiated jets peak around \RL$=0.08$, while quark-initiated jets exhibit higher distribution values and are more concentrated in the small-\RL~region. Inclusive charged jets represent a mixture of the two, and their distribution closely resembles that of gluon-initiated jets. This observation also reveals that within this kinematic regime, the charged-hadron jet sample is predominantly dominated by gluon-initiated jets. The transition from gluon-initiated to quark-initiated jets produces an effect similar to what is achieved by increasing the jet $p_{\rm T}$.

{\textcolor{blue}{{\bf \textit{different jet quenching mechanism}}~-~} In Fig.~\ref{fig:qgratio}, within the same kinematic regime, the jet quenching patterns for different flavor-initiated jets are investigated individually.  First of all, we present the A+A/p+p ratios of the $\Sigma_{\text{EEC}}$ distributions for quark-initiated jets and gluon-initiated jets in the upper and lower panels, respectively. Below \RL$=0.15$, a slight and gradual suppression is observed for quark-initiated jets in the upper panel, while beyond \RL$=0.15$, the ratio exhibits an increasing behavior with rising \RL. The enhancement observed in the large-\RL~region reveals a distinct diffusion effect induced by jet quenching for quark-initiated jets. However, the suppression seen in the small-\RL~region indicates that the selection bias effect is not manifested. In the lower panel, we observe a significant enhancement for gluon-initiated jets in the small-\RL~region below $0.07$. This enhancement becomes more pronounced as the \RL decreases. Additionally, a substantial increase is also seen in the large-\RL~region beyond 0.4, where the ratio rises monotonically with increasing \RL.  

Fig.~\ref{fig:baseline} indicates that jets with higher transverse momentum tend to populate the distribution in smaller-\RL~region. The enhancement observed in the small-\RL~region of this figure likely originates from contributions of higher-\(p_T\) jets before significant energy loss occurs. Notably, quark- and gluon-initiated jets exhibit markedly distinct responses to this mechanism. This difference may partly be explained by examining Fig.~\ref{fig:quarkgluon}: quark-initiated jets predominantly distribute in intrinsically smaller \RL~regions, whereas gluon-initiated jets favor larger \RL~regions. Furthermore, in the large-\RL~region, the observed enhancement is smaller for gluon jets than for quark jets. These characteristic differences collectively constitute the discriminatory power of the EEC observable in distinguishing the jet quenching signatures of quark-initiated versus gluon-initiated jets.
\begin{figure}[!htb]
\centering
\includegraphics[scale=0.45]{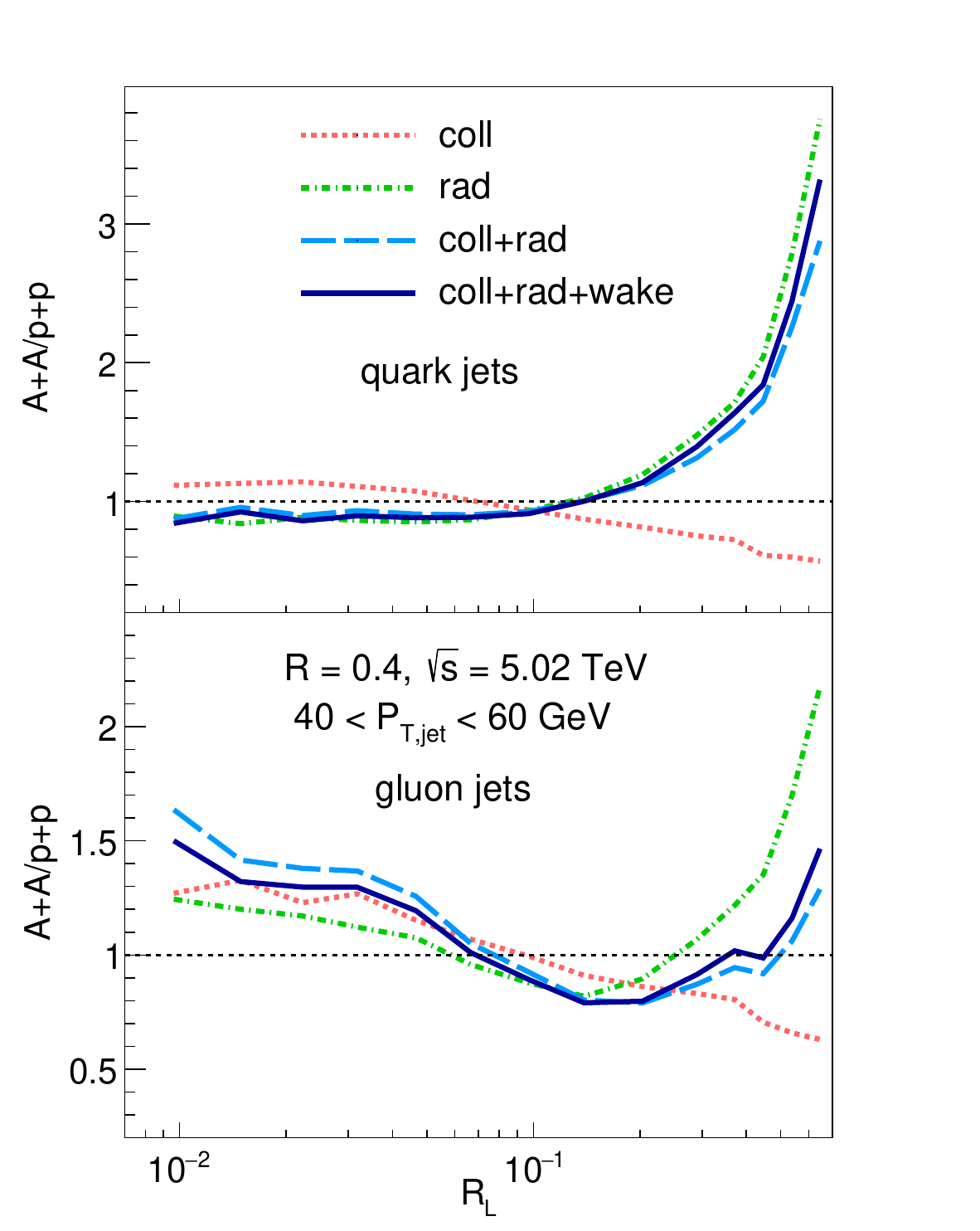}
\caption{ A+A/p+p ratios of the $\Sigma_{\text{EEC}}$ distributions for quark-initiated jets (upper panel) and gluon-initiated jets (bottom panel) separately calculated in the four jet quenching scenarios plotted in the same manner and the same kinetic region as in Fig.~\ref{fig:aappall}.} 
\label{fig:qgratio}%{}
\end{figure}

Simultaneously, we also computed the isolated impact of different jet quenching mechanisms on their respective A+A/p+p ratios, shown alongside the curve representing the overall consideration of the jet quenching mechanisms, following the methodology employed in Fig.~\ref{fig:aappall}. For the pure elastic energy loss mechanism, we find that regardless of whether for gluon-initiated or quark-initiated jets, it consistently leads to enhancement in the small-\RL~region and suppression in the large-\RL~region. This tends to increase their relative weighting at small \RL. However, the effect is less pronounced for quark-initiated jets compared to gluon-initiated jets, noting that initially, quark jets already exhibit a stronger tendency to populate the small-\RL~region than gluon jets. In contrast, the radiative energy loss mechanism exhibits significantly greater differences between quark-jets and gluon-jets than elastic energy loss. For quark-initiated jets, radiative energy loss manifests as suppression in the small-\RL~region and enhancement in the large-\RL~region. Its behavior almost entirely dictates the overall A+A/p+p ratio pattern, rendering the contribution from elastic energy loss negligible. The medium response effect provides only a minor enhancing contribution in the large \RL~region. For gluon-initiated jets, radiative energy loss manifests as an enhancement in both the small and large \RL~regions. Overall, radiative energy loss still dominates the A+A/p+p ratio for gluon jets. However, its dominance is more substantially diminished by the opposing effects of elastic energy loss compared to the case in quark-initiated jets, where it holds a stronger position.

{\textcolor{blue}{{\bf \textit{selection bias effect}}~-~} Through the above analysis and comparisons, we recognize the critical need to further characterize the selection bias effects experienced separately by quark- and gluon-initiated jets, moving beyond intuitive interpretations based solely on the $p_{\rm T}$-dependence observed in Fig.~\ref{fig:baseline}. We employ the following methodology: Jets with reconstructed transverse momentum in the $ 40- 60$~GeV range in A+A collisions are categorized into two classes: \textit{Falldown}, jets originating from partons that, prior to energy loss, would have produced jets with $p_{\rm T} > 60$~GeV in the corresponding p+p collisions.  \textit{Survival}, jets originating from partons that, before energy loss, would have produced jets with $p_{\rm T}$ in the $40- 60$~GeV range in the corresponding p+p collisions. In Fig.~\ref{fig:double-ratio}, we comparatively present the A+A/p+p ratios of the EEC distributions for both quark-initiated and gluon-initiated jets, separately showing the contributions originating from the  \textit{Falldown} and  \textit{Survival} components for each flavor. 

We find that the energy loss patterns manifested by the \textit{Falldown} and \textit{Survival} components are dynamically analogous. The characteristic differences in energy loss signatures fundamentally originate from the flavor of the initial partons. Comparatively, the \textit{Falldown} contribution consistently exhibits a stronger enhancing effect in the small-\RL~region while producing relatively weaker diffusion in the large-\RL~region. The combined effect observed in Fig. 3 arises from the mixture of these \textit{Falldown} and \textit{Survival} components. Examining the \textit{Survival} component in isolation reveals that in the small-\RL~region, even without the amplified selection bias effect intrinsic to the \textit{Falldown} population, gluon-initiated jets inherently generate a slight enhancement due to their characteristic energy loss pattern. The selection bias effect further augments this enhancement. Simultaneously, we note that the enhancing effect of selection bias in this region remains insufficient to counteract the suppressing effect of quark jet energy loss. This leads to the empirical observation that gluon jets appear more susceptible to selection bias. The underlying physics, however, indicates that both quark and gluon jets experience selection bias. The observed difference primarily arises from the flavor-dependent characteristics of energy loss imprinted on the EEC observable.

%There is a reminder that a competition between the energy loss effect and the selection bias effect is taking place for both pure quark and gluon jets, but why the difference? We can roughly imagine that the intensity of the energy loss plays an important role in the selection bias effect, which is responsible for the EEC enhancement at smaller $R_{\rm L}$. We also know that gluons tend to lose larger energy than quarks.  Since there is obvious enhancement at smaller $R_{\rm L}$ for gluon jets, it seems the energy loss effect of quark jets overwhelms its selection bias effect. However, the reasoning comes to a paradox: larger energy loss leads to a broadening of particle pair angular and strong selection bias. however, quark loses less energy than gluon but suffers stronger broadening.

\begin{figure}[!htb]
\centering
\includegraphics[scale=0.45]{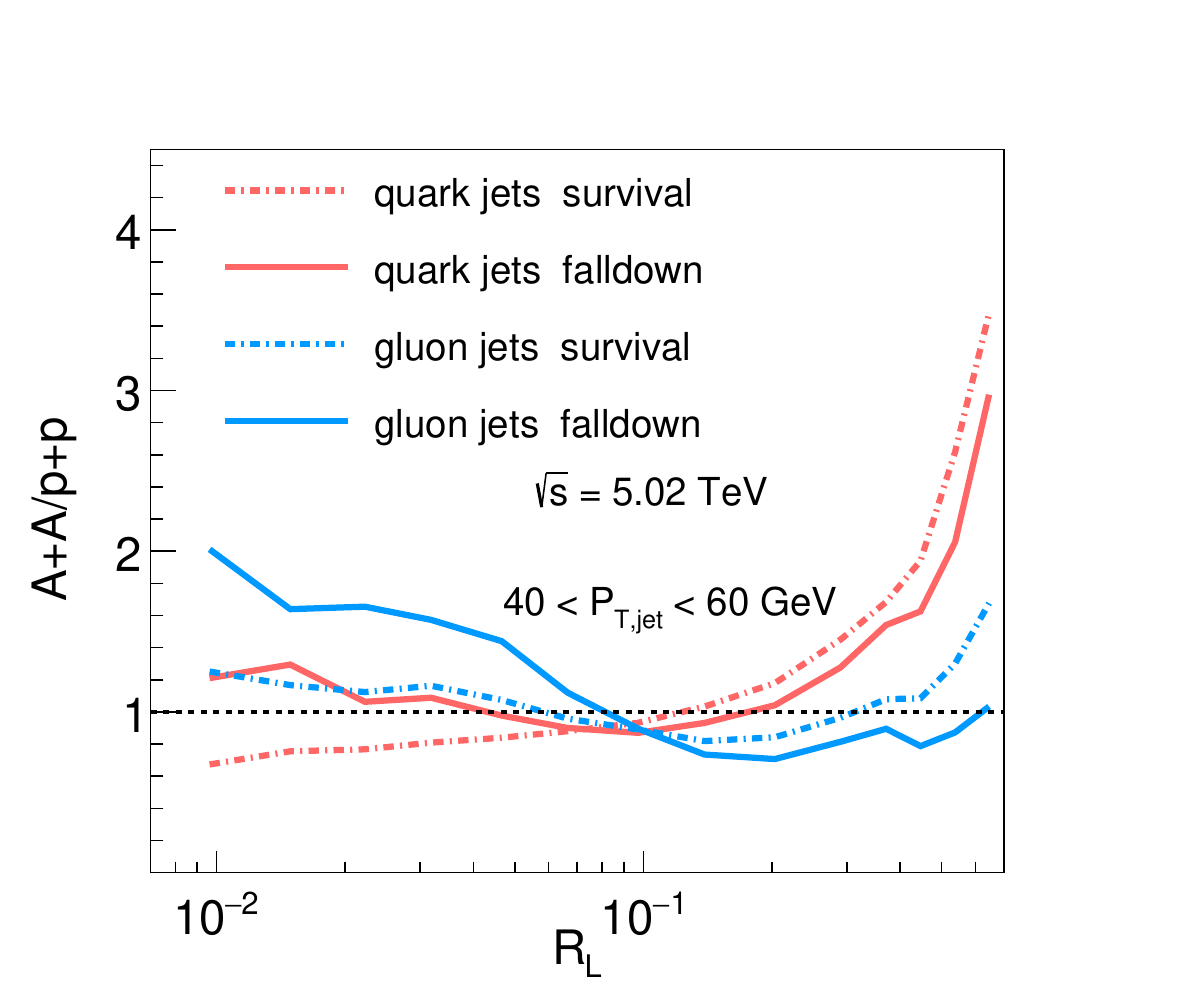}
\caption{A+A/p+p ratios for EEC distributions as a function of $R_{\rm L}$ for quark- and gluon-initiated jets with two categories: \textit{Survival} and \textit{Falldown} in central~($0-10\%$) Pb+Pb collisions at $\sqrt{s}=5.02$~ TeV in jet $p_{\rm T}$ interval $40-60$~GeV.} 
\label{fig:double-ratio}%{}
\end{figure}

{\textcolor{blue}{{\bf \textit{in-jet finer substructure of EEC}}~-~} To gain further insight into the aforementioned differences in jet quenching effects between quark- and gluon-initiated jets through the substructure details revealed by the EEC, we present the following in Fig.~\ref{fig:rjet}. Upper panels: the A+A/p+p ratios of the normalized paired-particle count \RL~distributions (normalized per jet) for quark-initiated jets (left) and gluon-initiated jets (right). Lower panels: the A+A/p+p ratios of the $\langle\mathrm{weight}\rangle$s \RL~distributions for quark-initiated jets (left) and gluon-initiated jets (right). Additionally, results showing the isolated effects of different jet quenching mechanisms are overlaid in each panel. 

Let's begin with the A+A/p+p ratios of the normalized paired-particle count \RL~distributions. For quark-initiated jets: The angular distribution of paired particles diffuses toward larger \RL~values. Elastic energy loss suppresses this trend by inducing suppression at large \RL, but the medium response effect compensates for this suppression. Consequently, the net effect nearly matches that of radiative energy loss. For gluon-initiated jets: The physical behavior pattern is analogous, although the diffusion toward larger \RL~occurs to a lesser extent than in quark jets. 

Turning now to the A+A/p+p ratios of the $\langle\mathrm{weight}\rangle$ \RL~distributions: for quark-initiated jets, modifications to the $\langle\mathrm{weight}\rangle$ are relatively mild overall. A slight enhancement only begins to emerge beyond \RL$=0.3$. Furthermore, the entire modification remains dominated by radiative energy loss, even though the elastic energy loss mechanism exhibits opposing behavior, consistent with the indication from the upper panel. Similarly, we observe that the medium response effect provides a significant suppressing contribution in the large-\RL~region, directly counteracting the enhancement induced by radiative energy loss. Consequently, this results in an overall gentle modification, suggesting that the $\langle\mathrm{weight}\rangle$ distribution experiences no significant shift across the entire \RL~range. For gluon-initiated jets, the shift of the $\langle\mathrm{weight}\rangle$ towards concentration in the small-\RL~region is markedly pronounced. Both radiative and elastic energy loss contribute in the same direction (towards this shift), and the medium response effect further suppresses the distribution in the large-\RL~region.

\begin{figure}[!htb]
\centering
\includegraphics[scale=0.45]{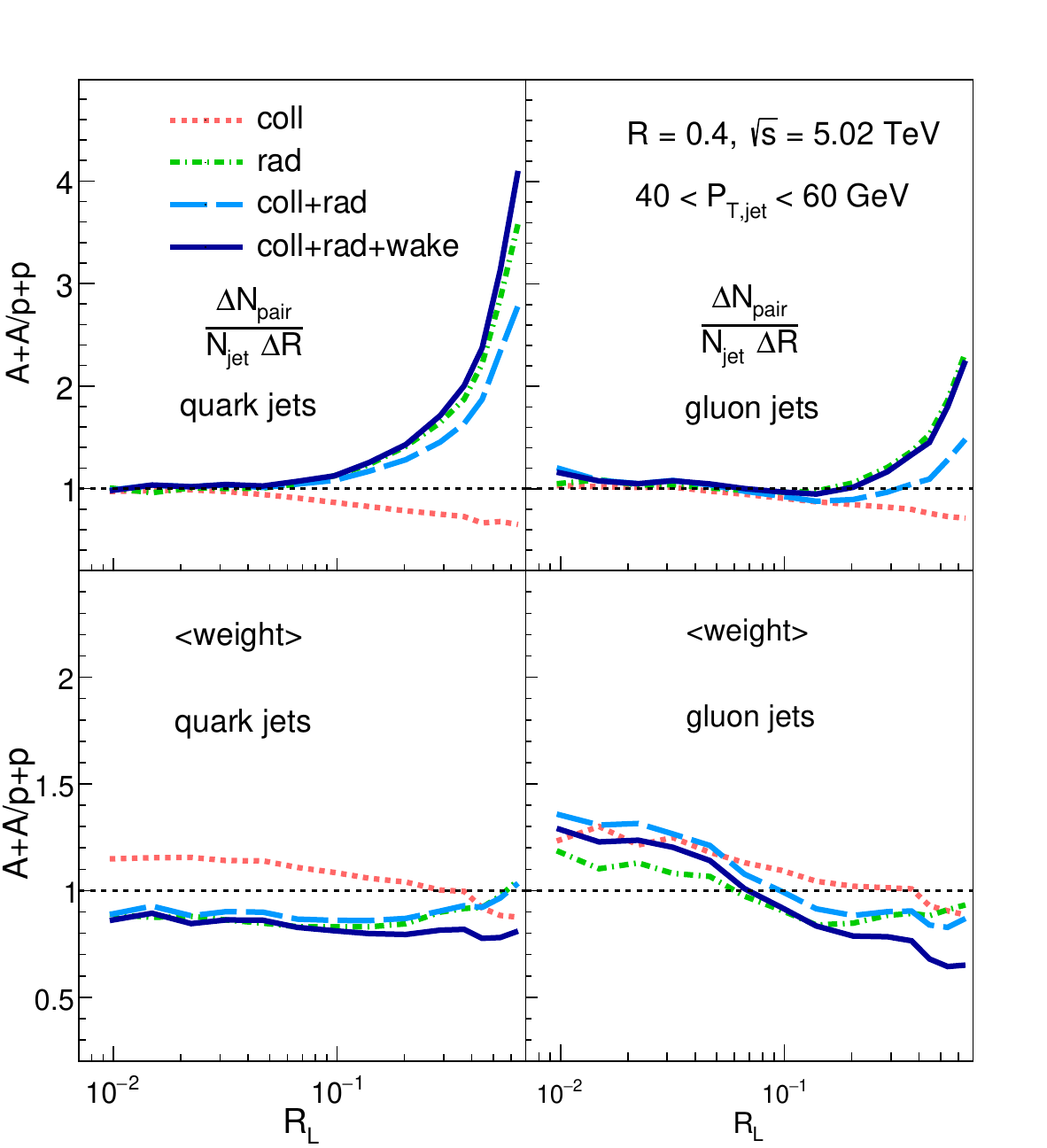}
\caption{The A+A/p+p ratios of the number of jets normalized paired-particle count \RL~distributions for quark-initiated jets (left) and gluon-initiated jets (right) are plotted in the upper panels; The A+A/p+p ratios of the $\langle\mathrm{weight}\rangle$s \RL~distributions for quark-initiated jets (left) and gluon-initiated jets (right) are plotted in the lower panels. The jets are with a jet size of $R=0.4$ in the jet transverse momentum interval of $40 \ {\rm GeV} <p_{T, \rm jet}<60 $~GeV produced in $\rm Pb+Pb$ collisions at $\sqrt{s}=5.02$~TeV.}
\label{fig:rjet}%{}
\end{figure}

Building on the two refined aspects of jet substructure quenching effects, we now analyze the results in Fig.~\ref{fig:qgratio}, which reflect their combined impact. For quark-initiated jets, elastic energy loss contributes negligibly. Both radiative energy loss and medium response increase the number of jet constituents. This significantly enhances the paired-particle count \RL~distribution while substantially suppressing the $\langle\mathrm{weight}\rangle$. In the small-\RL~region, minimal modification to paired-particle counts dominates. The suppression of $\langle\mathrm{weight}\rangle$ thus causes an overall suppression in the total EEC A+A/p+p ratio. In the large-\RL~region, the strong enhancement in paired-particle counts is partially counteracted by suppressed $\langle\mathrm{weight}\rangle$. Nevertheless, a significant net enhancement persists. For gluon-initiated jets, the enhancement in the paired-particle count \RL~distribution is weaker than in quark jets. The critical difference lies in modifications to the $\langle\mathrm{weight}\rangle$ \RL~distribution: in the small \RL~region, an enhancement is observed; in the large \RL~region, pronounced suppression is observed. Consequently, enhanced $\langle\mathrm{weight}\rangle$ drives an overall enhancement in the total EEC A+A/p+p ratio in the small-\RL~region, while suppression of $\langle\mathrm{weight}\rangle$ further diminishes the enhancement of the paired-particle counts in the large-\RL~region.
% \section{Discussion and conclusions}\label{sec:5}

%{\textcolor{blue}{{\bf \textit{Attribution and summary}}~-~ }

\section{Attribution and summary}\label{sec:4}

We have analyzed the differences in the A+A/p+p \RL distributions between quark and gluon jets through four distinct dimensions: initial production mechanisms, variations in jet quenching mechanisms, internal substructure details revealed by EEC, and differential susceptibility to selection bias. We now investigate the fundamental origins of the flavor-dependent jet quenching effects observed in the EEC.

We therefore compute the A+A/p+p ratios of the EEC distributions under two counterfactual scenarios: pure gluon jets traversing the medium with quark-like energy loss characteristics, pure quark jets subjected to gluon-like energy loss dynamics. Results are presented in Fig. \ref{fig:qasg}. These variations arise not only from differences in the interaction strength between quarks/gluons and the hot and dense medium, but also from distinct radiation patterns of gluons in radiative energy loss. We quantify this impact by highlighting the variation range between the nominal A+A/p+p ratio and our counterfactual ratio through shaded bands in the figure. For gluon jets with suppressed energy loss, we observe the expected attenuation of both energy loss and selection bias effects. However, enhancements persist across both small and large $R_{\rm L}$ regions. Conversely, enhancing quark energy loss to gluon-like levels dramatically amplifies the EEC distribution in A+A collisions at large $R_{\rm L}$ while preserving the characteristic suppression at small $R_{\rm L}$. 

\begin{figure}[!htb]
\centering
\includegraphics[scale=0.45]{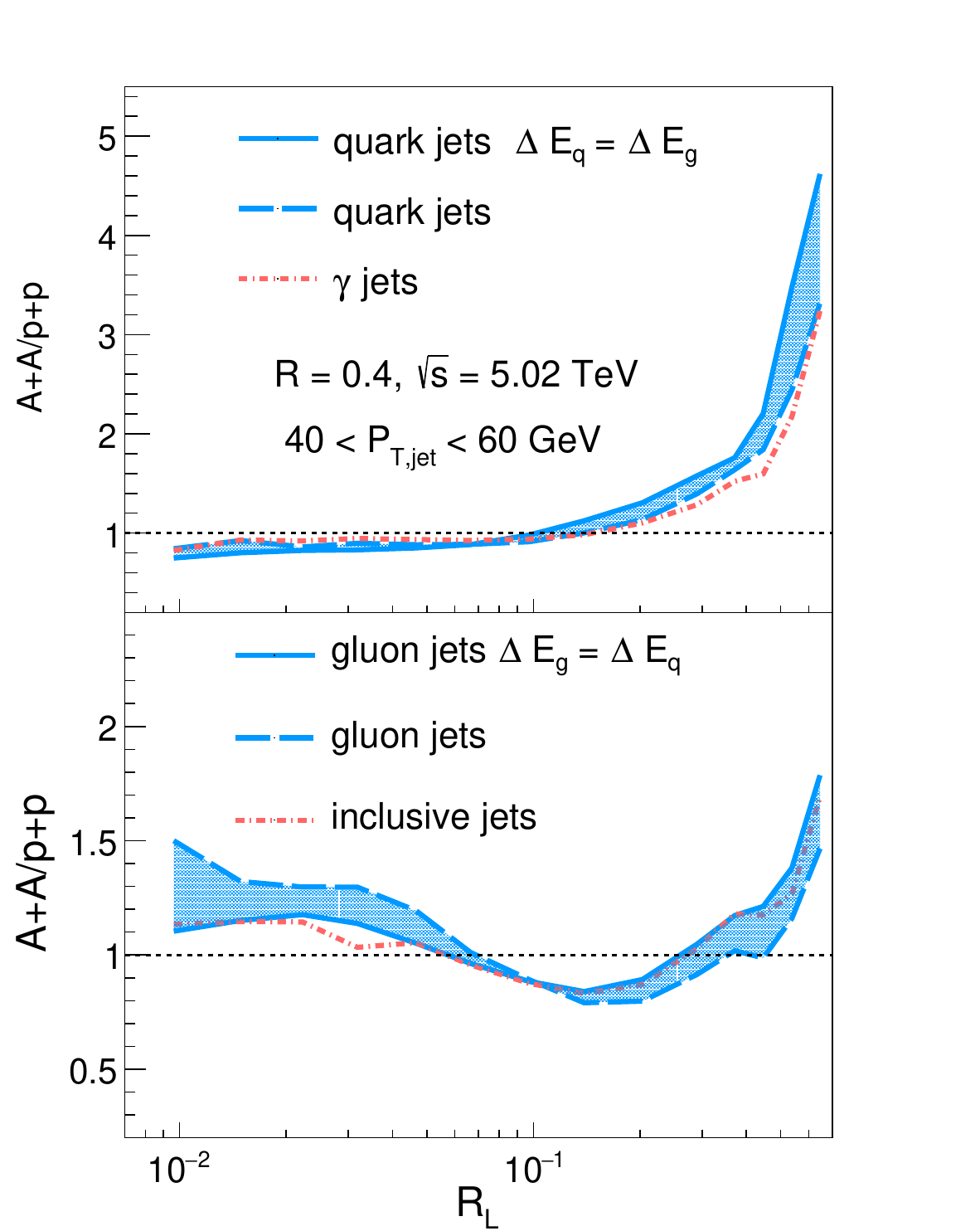}
\caption{Counterfactual analysis of flavor-dependent jet quenching via EEC distributions. (upper) Quark jets with energy loss equivalent to gluon jets ($\Delta E_q = \Delta E_g$), compared to $\gamma$-tagged quark jets under standard quenching. (lower) Gluon jets with energy loss equivalent to quark jets ($\Delta E_g = \Delta E_q$), compared to inclusive jets under standard quenching. All jets reconstructed with $R=0.4$ in central $(0-10\%)$ Pb+Pb collisions at $\sqrt{s_{NN}}=5.02$ TeV with transverse momentum $40 < p_{\rm T,jet} < 60$ GeV. Shaded bands indicate the variation range between nominal and counterfactual A+A/p+p ratios.} %Í¼Ìâ
\label{fig:qasg}%{}
\end{figure}

Consequently, the characteristic differences in A+A/p+p ratios of EEC distributions between quark- and gluon-initiated jets primarily stem from pre-quenching distribution disparities, not from flavor-dependent energy loss mechanisms, and crucially not from divergent gluon radiation patterns during radiative energy loss. 

To experimentally observe and utilize this flavor-dependent difference, we propose photon-tagged jets as proxies for quark-initiated jets and inclusive charged-hadron jets within the same kinematic regime as proxies for gluon-initiated jets. We comparatively present the photon-tagged jets versus pure quark-initiated jets in the upper panel. The A+A/p+p ratios of EEC distributions for both jet types are compared within the same kinematic window. The photon-tagged jet results show notably closer alignment with unmodified quark-initiated jets. The inclusive charged-hadron jets are compared to pure gluon-initiated jets in the lower panel.  The results for inclusive charged-hadron jets align closely with those of gluon-initiated jets, which have been artificially modified to undergo quark-like energy loss.

In summary, through PYTHIA simulations validated against ALICE p+p data and an improved SHELL model incorporating collisional/radiative energy loss plus medium response, we systematically quantify flavor-dependent jet quenching via energy-energy correlators (EEC) in $\sqrt{s_{NN}}=5.02$ TeV Pb+Pb collisions. Key findings include that the flavor-dependent quenching signatures in this study are characterized by pure quark jets exhibiting a strong enhancement at $R_L > 0.2$, and gluon jets displaying a bimodal enhancement at both small ($R_L < 0.07$) and large ($R_L > 0.4$) angular scales. Dual-shift decomposition is illustrated in the A+A/p+p ratio of the EEC, where shifts toward large \RL~ ($>0.3$) primarily stem from energy loss effects. Shifts toward small \RL ($<0.06$) are not fully attributable to selection bias. Counterfactual analysis reveals that intrinsic gluon jet enhancement is also evident in small-\RL regions. The aspect of the substructure-resolved discrimination is that the quark jets suffer global suppression of $\langle\mathrm{weight}\rangle(R_L)$ and gluon jets suffer concentration of $\langle\mathrm{weight}\rangle$ toward small~\RL. From the mechanism decomposition point of view, elastic energy loss concentrates $\langle\mathrm{weight}\rangle(R_L)$ toward small~\RL, radiative energy loss dominates quark jet modifications, and medium response amplifies large \RL~enhancement via soft hadron production. These results indicate that flavor dependence in EEC modifications is dominantly driven by intrinsic gluon/quark-initiated jet distribution structure differences rather than medium-induced mechanisms. At the end of the manuscript, we propose photon-tagged jets as proxies for quarks and inclusive charged-hadron jets as proxies for gluons. $\gamma$-tagged jets reproduce the pure quark jet quenching signatures, and inclusive jets exhibit gluon-like behavior when subjected to quark-like energy loss. These results establish the EEC as a precision probe of color-charge-dependent jet-medium interactions, providing new constraints for $\hat{q}$ extraction and QGP tomography. Meanwhile, the dominant role of initial jet structure suggests future analyses must account for pre-quenching flavor asymmetries.

{\bf Acknowledgments:}  This research is supported by
the Guangdong Major Project of Basic and Applied
Basic Research No. 2020B0301030008, and the National Natural Science Foundation of China with Project
Nos. 11935007 and 12035007. Ke-Ming Shen is supported by the Doctoral Research of ECUT (Nos. DHBK2019211).

\vspace*{-.6cm}

%\bibliographystyle{unsrt}
%\bibliography{ref}

\end{document}